%% file: mastr.tex
\documentclass[acmtog,screen,nonacm,review=false,timestamp=false]{acmart}
\usepackage{tikz}

\newenvironment{datamaterial}%
{ \vspace{-0.15cm}%
    \small\noindent{\bfseries Availability of Data and Material:}\par%
    \noindent\ignorespaces}%
{ \par\noindent%
\ignorespacesafterend }%

\AtBeginDocument{%
  }

\begin{document}

\title{Monitoring Germany's Core Energy System Dataset: A Data Quality Analysis of the Marktstammdatenregister}
\author{Florian Kotthoff}
\email{florian.kotthoff@offis.de}
\orcid{0000-0003-3666-6122}
\affiliation{%
  \institution{OFFIS Institute for Information Technology}
  \streetaddress{Escherweg 2}
  \city{Oldenburg}
  \state{Lower Saxony}
  \country{Germany}
  \postcode{26121}
}
\affiliation{%
  \institution{fortiss}
  \streetaddress{Guerickestraße 25}
  \city{Munich}
  \state{Bavaria}
  \country{Germany}
  \postcode{80805}
}
\author{Christoph Muschner}
\orcid{0000-0001-8144-5260}
\affiliation{%
  \institution{Reiner Lemoine Institut gGmbH}
  \streetaddress{Rudower Chaussee 12}
  \city{Berlin}
  \state{Berlin}
  \country{Germany}
  \postcode{12489}
}
\author{Deniz Tepe}
\orcid{0000-0002-7605-0173}
\affiliation{%
  \institution{fortiss, Research Institute of the Free State of Bavaria}
  \streetaddress{Guerickestraße 25}
  \city{Munich}
  \state{Bavaria}
  \country{Germany}
  \postcode{80805}
}
\author{Esther Vogt}
\affiliation{%
  \institution{Institute for Enterprise Systems, University of Mannheim}
  \streetaddress{Guerickestraße 25}
  \city{Mannheim}
  \state{Baden-Württemberg}
  \country{Germany}
  \postcode{68131}
}
\author{Ludwig Hülk}
\orcid{0000-0003-4655-2321}
\affiliation{%
  \institution{Reiner Lemoine Institut gGmbH}
  \streetaddress{Rudower Chaussee 12}
  \city{Berlin}
  \state{Berlin}
  \country{Germany}
  \postcode{12489}
}

\begin{abstract}
The energy system in Germany consists of a large number of distributed facilities, including millions of PV plants, wind turbines, and biomass plants. 
To understand and manage this system efficiently, accurate and reliable information about all facilities is essential.
In Germany, the Marktstammdatenregister (MaStR) serves as a central registry for units of the energy system.
The reliability of this data is critical for the registry's usefulness, but few validation studies have been published.

In this work we provide a review of existing literature that relies on data from the MaStR and thereby show the registry's importance.
We then build a data and testing pipeline for relevant data of the registry, with a focus on the two aspects of facility's location and size.  
All test results are published online in a reproducible workflow.
Hence, this work contributes to a reliable data foundation for the German energy system and starts an open validation process of the Marktstammdatenregister from an academic perspective.

\end{abstract}




\maketitle

\begin{datamaterial}

The examined dataset version can be found on zenodo \cite{hulk2022_zenodo}. Downloading and Preprocessing was done with the python package open-mastr \cite{Hulk_open-MaStR_2023}. The data pipeline and dashboard code can be found in our github repository \cite{githubrepo}.

\end{datamaterial}

\section{Introduction} \label{sec:intro}
The current energy system is undergoing an important transformation, with a shift from a few large and mainly fossil power plants to an increasing number of small renewable power units. The wind and solar technologies, in particular, experience rapid growth rates, with 35.000 wind turbines and 3.9 million photovoltaic (PV) systems producing electricity in Germany as of early 2024 \cite{Bundesnetzagentur2019_Marktstammdatenregister}. To understand and control this diverse and complex energy system, detailed and reliable data of all its components is essential. Such an accessible and up-to-date data source is a key enabler for the digitalization within the energy system \cite{Hack2021}. 

For this reason, the German legislator has created a large and central power plant registry, the \textit{Marktstammdatenregister}, or short, MaStR. It was released in 2019, is updated on a daily basis, and contains more than 22 million entries about electricity and gas producers, electricity and gas consumers, storages, grids, and actors from the energy market. The data is provided by the owners of the power plants themselves. In this sense, creating this dataset was very expensive and time-consuming, as millions of owners had to be contacted, needed to register on a website and enter their information. Validation and reliability assessment are of great importance, as the data is intended to play a central role in controlling the future German energy system and is based on the contribution and information of millions of individuals.

Considering the registry's increasing importance, the quality of the dataset should be monitored and discussed. In this paper, we start this work by building a data pipeline together with automated data tests for the MaStR. 
Since we do not have the rights to change the content of the dataset, we focus on finding and visualizing data errors that appear in the MaStR. By this, we want to answer the research question of \textit{how reliable is the data of the Marktstammdatenregister}, with a special focus on the units' location and size.

The core contributions of this work are the following: We are the first work that reviews the usage of the MaStR dataset in research to identify its importance and we contribute with a first step to the urgently needed validation of this central German energy system dataset. Our code, together with all test results and visualization dashboards, are published openly and can be utilized by other MaStR users.

This paper is structured as follows: In section \ref{sec:mastr}, we first describe the MaStR dataset. Then, in section \ref{sec:Literature}, we review and categorize the usage of the MaStR dataset in existing literature. In section \ref{sec:method}, we describe the data pipeline and the data tests. In section \ref{sec:results}, we analyze the test results, discuss them in section \ref{sec:discu} and draw our conclusion in section \ref{sec:conclu}. 

\section{The dataset - Marktstammdatenregister} \label{sec:mastr}
The MaStR is operated by the German Federal Network Agency (ge: Bundesnetzagentur, short: BNetzA) since January 31, 2019 as a central online database of the German energy system. Owners of electricity or gas generating units are obliged to report master data on themselves and their units. Additionally, units consuming large amounts of electricity have to be registered if they are connected to at least a high-voltage electricity grid.
Considering the type of information collected, the MaStR only includes master data like installed capacity or location information. It does not provide transaction data like actual production amounts or storage levels ~\cite{Bundesnetzagentur2018_Gesamtkonzept}. Following the introduction of the MaStR, all information had to be provided manually by plant owners using a registration website. A transition period of two years until 01/31/2021 was granted for existing plants, but plants which were commissioned since then have to be registered within one month after commissioning~\cite{BundesnetzagenturElektrizitaet2022_Monitoringbericht}. 

Most information on units is openly accessible. The data is published under an open data license, the Data licence Germany – attribution – version 2.0 (DL-DE-BY-2.0) and can be downloaded, used and republished with no restriction if proper attribution to the Bundesnetzagentur is given.
For units with a net capacity of up to 30 kW, some location information is restricted from publication. This applies to street name, house number, parcel designation and exact coordinates of units. The most granular location information accessible for all units is the postal code or the municipality. The MaStR dataset referenced in this work was downloaded at the 2024-03-12 from the official website of the BNetzA using the \textit{open-mastr} python package \cite{Hulk_open-MaStR_2023}. We have developed and published this package to provide a simplified and automated download, processing and storage in a local database. Additionally, we published the MaStR dataset from this date on zenodo \cite{hulk2022_zenodo}.

Regarding the validation process of the data, information on units, plants and corresponding owners has to be verified by the Distribution System Operator (DSO). In \cite{BundesamtJustiz_MaStR}, it is specified that the capacity and address, but not the coordinates of a power unit have to be validated by the DSO. Independent of the mandatory validation by the DSO, the BNetzA runs checks on the dataset as well. According to~\cite{BundesnetzagenturElektrizitaet2022_Monitoringbericht}, they recently focused on dimensioning errors of capacities and location information of wind turbines. For units with a net capacity of more than 10~MW, the BNetzA checks all reported information. Obvious errors are fixed immediately and reported to the plant owners who may object. On average, the BNetzA fixes about 600 errors per month for the whole registry.

Besides this work, several other publications validated the MaStR dataset. Manske et al. developed a pipeline to extend and validate the entries for PV systems, wind turbines, biomass plants, and hydropower plants in the MaStR \cite{Manske2022_Geo}. Mayer et al. applied a convolutional neural network (CNN) model to recognize PV systems from aerial images and provide an alternative PV dataset for one German state \cite{Mayer2020_DeepSolar}. Expanding on this, they took 3D building data into consideration and focused on the orientation and tilt angle of rooftop PV systems. By comparing the MaStR to their own dataset, they pointed out the discrepancies between both datasets \cite{Mayer2022_3D}. Schulz et al. mentioned the deficits in the MaStR and used a similar object recognition framework for PV, wind, biomass systems as well as solar thermal units. They aimed to supplement the official dataset and demonstrated the significance of unregistered units as case studies in two districts \cite{Schulz2021_DetEEktor}.

\section{Literature review} \label{sec:Literature}
To understand the current relevance of the MaStR, we conducted a review of existing literature. 
Therefore, we used the search query "Marktstammdatenregister" and searched within the science databases of \textit{Scopus}, \textit{Web of Science}, \textit{arxiv}, \textit{IEEExplore}, \textit{MDPI}, \textit{SpringerLink}, \textit{ScienceDirect}, and \textit{GoogleScholar}. The result was a total of 532 documents by $1^{\text{st}}$ April 2024. After removing duplicates and documents that were not peer reviewed or published within a scientific journal, 99 papers remained that contain the word "Markstammdatenregister". From these 99 works, a total of 83 works actually used the dataset and not only referred to it. These works, that build upon the MaStR dataset, are now described and categorized in further detail.

\begin{figure}[htb]
    \centering
    \includegraphics[width=\columnwidth]{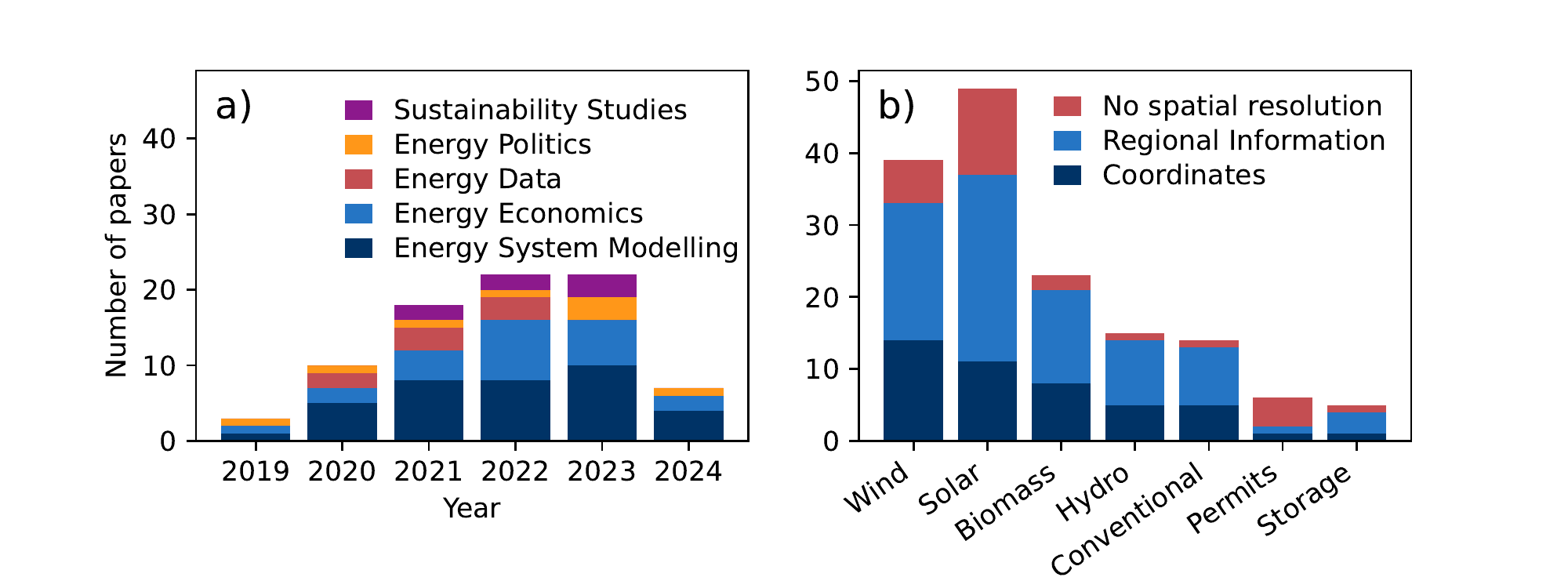}
    \caption{Importance of MaStR in literature: In a), the number of published papers for the five identified research domains is plotted over the publication year (up to $1^{\text{st}}$ April 2024). In b), the number of papers that use different tables from the MaStR is plotted. The papers are further subdivided according to the used location information, where \textit{regional information} represents the use of aggregated data on zip-code, district, or state level.}
    \label{fig:LiteratureReview}
    \Description{Figure a): Four bar charts showning the number of new papers using the MaStR from 2019 to 2022. In 2019, four papers were published. This number increases roughly linear up to 24 paper in the year 2022. The papers are further subdivided into the categories they fall into, with Energy System Analytics and Energy Economics having the largest share. In Figure b) seven bar charts show the number of papers that use information of different tables. Each table (wind, solar, biomass, hydro, conventional, permits, storage) have one bar. Solar has the largest number of papers (about 35), wind has a little less than 30. Storages has the smallest number with roughly 5.}
\end{figure}

From the 83 papers that use data from the MaStR, 48 papers are Open-Access publications. Since it is a relatively new dataset, the first publications using the dataset appeared in 2019, followed by an increased number of publications in the proceeding years. To get an idea which scientific fields use the MaStR, we assigned one research domain to each paper: \textit{Sustainability Studies} if the paper focused on environmental impact, \textit{Energy Politics} if the paper focused on energy politics and policy recommendations, \textit{Energy Data} if the paper focused on creating or evaluating datasets, \textit{Energy Economics} if the paper focused on economical aspects, and \textit{Energy System Analysis} if the paper focused on modelling and analysing the energy system. In Figure \ref{fig:LiteratureReview}a we see that the dataset was most frequently used in the domains \textit{Energy System Analysis} and \textit{Energy Economics}. In Table \ref{tab:LiteratureReview} we list all identified research domains together with the research topics that utilize data from the MaStR dataset. 

\begin{table*}[h!]
\caption{Description of the different domains and research questions where the MaStR dataset is used.}
\label{tab:LiteratureReview}
      \begin{tabular}{cl}
        \hline
        Research Domain  & Research Topic\\ 
        \hline
        Sustainability Studies & 
            \begin{tabular}[c]{@{}l@{}}
            - Land use of renewable energies \cite{Boehm2022_Land, Kyere2021_Spatio, Gassner2023, Christiansen2023}\\
            - Climate gas emissions due to power plants \cite{Eysholdt2022_model} \\
            - Waste quantification of decommissioned renewables \cite{Volk2021_Regional}
            \end{tabular}\\
        \hline
        Energy Politics & 
            \begin{tabular}[c]{@{}l@{}}
            - Local allocation of energy infrastructure \cite{Lueck2019_Wind, Fink2021_Der, Salomon2020_Minimum, Carattini2024, Schoenauer2023, Biehl2023} \\
            - Effects of \textit{Prosuming} \cite{Ouanes2022_Prosuming, Wege2023}
            \end{tabular}\\           
        \hline
        Energy Data  & 
            \begin{tabular}[c]{@{}l@{}}
            - Identifying renewable energy systems from aerial images \cite{Manske2022_Geo, Mayer2020_DeepSolar, Mayer2022_3D, Schulz2021_DetEEktor, Rausch2020_Enriched, Joerges2023} \\
            - Building time series datasets for energy system modelling \cite{Jesper2021_Annual, Bogensperger2022_Accelerating, Pflugfelder2024} \\
            - Developing ontologies for the energy domain \cite{Booshehri2021_Introducing}\\
            - Building datasets about energy cooperatives \cite{Wierling2023}\\
            \end{tabular}\\
        \hline
        Energy Economics &
            \begin{tabular}[c]{@{}l@{}}
            - Evaluating auction designs for renewable energies \cite{Kroeger2022_Discriminatory, Lineiro2021_Evaluating,  Lineiro2021_Lessons, Matthaeus2021_Renewable, Kitzing2022_Worth, Wrede2022_influence}\\
            - Evaluating costs of renewable energies \cite{Kraschewski2023} \\
            - Profitability and business models in the energy domain
             \cite{Wierling2021_Contribution, Wierling2022_Business, Engelhorn2022_Development, Loehr2022_Facing, Mueller2024, Stute2023} \\
            - Adoption of renewable energy technology
             \cite{Arnold2022_How, Lueck2019_Artificial, Johanning2022_PVactVal, Figgener2020_development, Galvin2022_Why, Ganz2024}  \\ 
            - Local energy markets \cite{Lueth2020_distributional, Kilthau2023}
            \end{tabular} \\   
        \hline
%
        \begin{tabular}[c]{@{}c@{}}
        Energy System \\ Analysis
        \end{tabular} & 
        \begin{tabular}[c]{@{}l@{}}
            - Flexibility options \cite{Doerre2021_Flexibility, Kondziella2021_Status, Stoessel2021_County, Schroeer2024} \\
            - Curtailed energy potentials \cite{Frysztacki2020_Modeling, Siddique2020_Assessment} \\
            - Repowering \cite{Grau2021_Sounding, Stetter2022_Hidden} \\
            - Analysis of models and methods \cite{Hoffmann2021_Typical, Ouwerkerk2022_Comparing, Unnewehr2022_value, Jaetz2023, EsmaeiliAliabadi2023} \\
            - Analysis of Electricity, Gas and Hydrogen Grids \cite{Meinecke2020_Simbencha, Mueller2022_Impact, Tran2019_Modelling, Triebs2021_Landscape, Weiss2021_Simulation, Wirtz2021_Modelling, Bartels2022_Integration, Buettner2024, Hobbie2024, Mertins2023, Weise2023, Lechner2023, Karalus2023} \\
            - Forecasting \cite{Schroeter2020_Substation, Li2023, Li2023a} \\
            - Potential analysis of renewable energies, storages, or electrolysis \cite{Kockel2022_Does, Bao2022_bottom, Jung2022_Development, Kuehnbach2020_How, Langenmayr2023, Walgern2023, Thraen2023, Lechl2023, Jung2023}
            \end{tabular} \\ 
        \hline
      \end{tabular}
\end{table*}

Since one contribution of our work is a validation of power unit locations within the MaStR, we also highlight the usage of location information in the literature. Therefore, each paper is categorized in Figure \ref{fig:LiteratureReview}b according to one of the three categories: \textit{No location information} if the location of power units was not used, \textit{Regional information} if either the state, the NUTS-region, or the municipality of a power unit was used, and \textit{Coordinates} if the actual longitude and latitude coordinates of a power unit were used within the publication. As a result, from the total of 83 papers we identified 24 papers which do not use location information, 33 papers which use regional information, and 26 papers which use coordinates. 

As a next step, we want to evaluate which MaStR tables are most frequently studied by researchers when using the dataset. Therefore, we identify the tables used in each publication. In Figure \ref{fig:LiteratureReview}b one can see that the two technologies of solar and wind receive the largest attention, followed by biomass, hydropower plants, Combined Heat and Power (CHP) plants and conventional plants (coal, gas, oil, and nuclear) \footnote{In contrast to further categorizations, a paper can have multiple tables assigned.}. Both the tables on storages and permits are used relatively rare, whereas permits are mainly used in the scope of energy economics to evaluate the performance of auctions or the average construction time of power units. For the four renewable technologies wind, solar, biomass and hydro, most works also use location information of the power units, either on a higher level (regional information) or the exact coordinates. 

\section{Method} \label{sec:method}
We now describe the whole procedure of obtaining, processing and testing the data. The details of the data pipeline are presented in section \ref{sec:method:DataPipeline}, whereas the data tests are presented in section \ref{sec:method:DataTest}.
\subsection{Data Pipeline} \label{sec:method:DataPipeline}
\begin{figure}[htb]
    \centering
    \includegraphics[width=0.9\columnwidth]{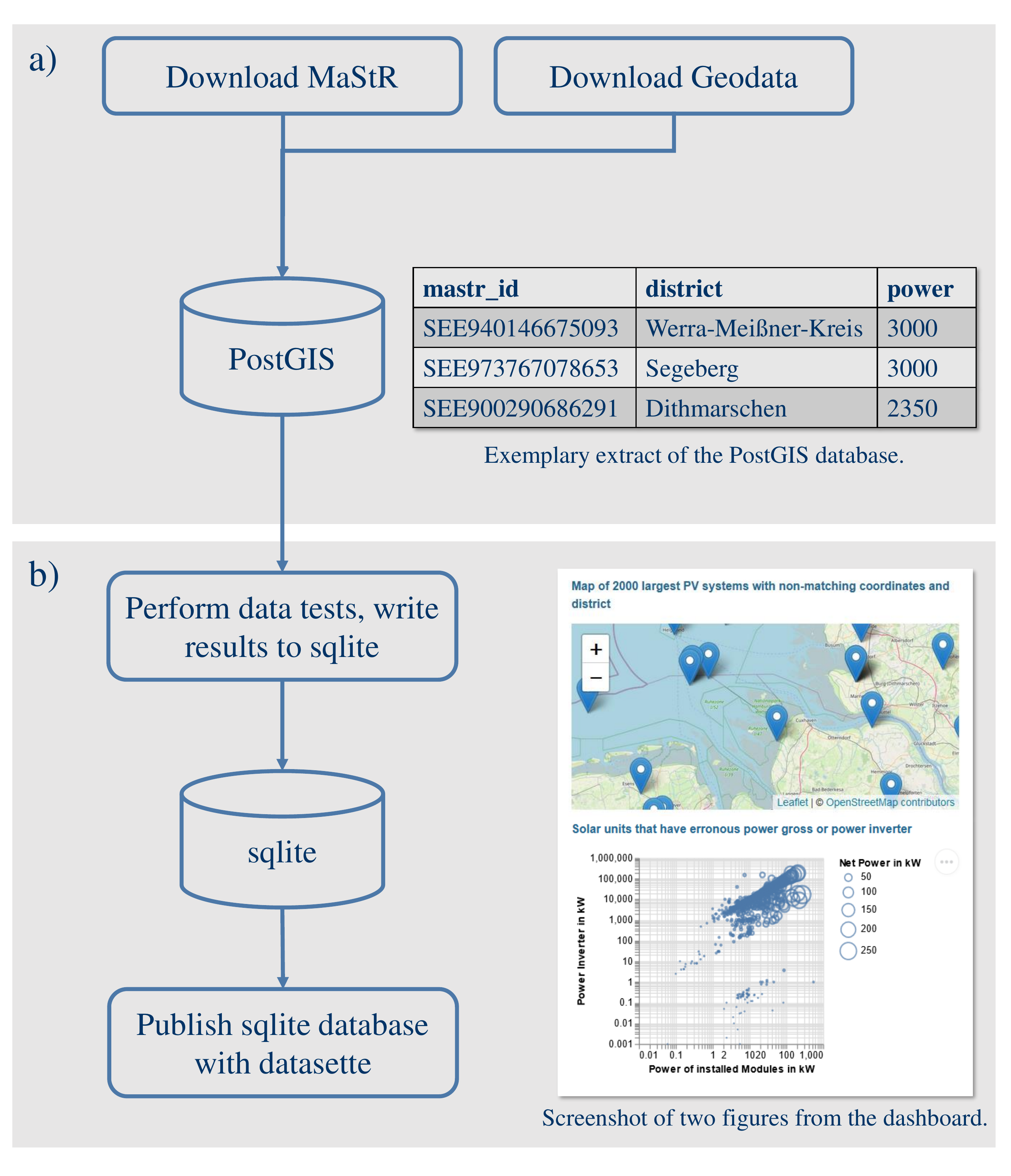}
    \caption{Automated pipeline for downloading, processing, testing, and vizualizing the MaStR dataset. In the first step (a), the required raw data is downloaded and written to a PostGIS database. Afterwards in (b), the data transformation and testing is performed using the framework dbt. All units that fail at least one test are written to an sqlite database. The sqlite database together with monitoring dashboards is then published using the framework datasette.}
    \label{fig:Pipeline}
    \Description{A woman and a girl in white dresses sit in an open car.}
\end{figure}
The data pipeline consists of multiple steps. First, the MaStR dataset is downloaded and parsed to a PostGIS database using the open-mastr python package \cite{Hulk_open-MaStR_2023}. Then, data of geoboundaries from districts and municipalities \cite{FACG2021_Administrative} is downloaded and saved to the same database. The geoboundaries are later used for testing unit locations. After all raw data is successfully loaded, it is transformed to a common data model using the software dbt \cite{labs2024_dbt}. The transformation consists of deleting unused information and renaming columns. After all tables are transformed to the same data model, the data tests are performed. Those tests are described in detail in Section \ref{sec:method:DataTest}. All units that do not pass the data tests are saved to a seperate sqlite database. The sqlite database is then published at \url{https://marktstammdaten.kotthoff.dev} using the framework datasette \cite{Willis2024_Datasette}. This publication consists of two parts: First, the whole database of units that fail one or more tests can be searched online. Subsets of the data can also be downloaded as CSV for further use. Second, dashboards are published that visualize the main error metrics we have defined to monitor the MaStR. The whole pipeline is shown in Figure \ref{fig:Pipeline}.

\subsection{Data Tests} \label{sec:method:DataTest}
Assuring high data quality is a very general problem, appearing both in science and industry. Pepino et al. \citep{pipino2002_data} define 16 dimensions of data quality. In the scope of monitoring the data quality of the MaStR dataset, two of those dimensions are of special interest: Completeness as "the extent to which data is not missing" and Free-of-Error as "the extent to which data is correct and reliable". Completeness can be defined as the fraction of entries that are not null and the total number of entries. The completeness of relevant parts of the MaStR is part of the the dataset description in the Appendix. 
Monitoring the "Free-of-Error" dimension however is complicated, as data correctness and reliability are closely coupled to the content of the data. To evaluate the "Free-of-Error" dimension, data unit testing comes into play. Data unit tests borrow the idea from software unit tests. They test small parts of the data to see if it behaves as expected. Several state-of-the-art tools implement data tests \cite{Gong_Great_Expectations, labs2024_dbt}. The data pipelining framework dbt used in this work also integrates data testing.

Technically, each data test is represented by one SQL query. The query

\begin{quote}
    SELECT * FROM wind WHERE mastr\_id IS NULL;
\end{quote}

represents the expectation that every MaStR unit has an ID. The test fails, if the query returns one or multiple rows. 

\begin{table*}
    \include{figures/table_testcases}
    \caption{Table of the different data tests. The column \textit{ID} is used to reference the tests in the text, \textit{description} gives a short summary of the test. A $\checkmark$ in the technology columns denote that the test is applied for these technologies.}
    \label{tab:tests}
\end{table*}

In table \ref{tab:tests} all implemented tests are shown. The tests belong to one of four categories, where numbers in [] denote the test IDs from table \ref{tab:tests}:

\begin{itemize}
    \item \textbf{Basic tests [1-5]:} For the MaStR dataset there are some baseline tests to ensure the data quality. We defined columns that are not allowed to have null values, checked that unit IDs are unique, that gross power and inverter power are larger then the net power, and that unit IDs, municipality IDs and zip codes match their respective regular expressions.
    \item \textbf{System size tests [6-9]:} The power of an electricity producing unit correlates with its system size. Hence we validated the rated power by comparing it with other quantities that represent system size. For storages and PV systems, the net power should be close to the power of the inverter. The power of PV systems should also correlate with the number of modules and, for ground-mounted PV systems, with the utilized area. For wind turbines, the rated power should correlate with the square of the rotor diameter.
    \item \textbf{Location tests [10,11]:} For units with a power larger 30 kWp, exact coordinates as well as an address is given. We checked if the coordinates lie in the given districts and municipalities, where the geoboundaries were taken from \cite{BundesamtKartographie_OpenData}. Since geoboundaries do not have an infinite resolution, we added a buffer zone of 1.5km to each boundary, so that units that lie on the boundary (or a maximum of 1.5km outside the district / municipality) still pass the test.
    \item \textbf{Technology specific tests [12-15]:} Some tests are specific to technologies and rely on domain knowledge. For each technology, we defined a range of years as accepted installation years. We chose 2030 as a maximum accepted year for planned installations. For wind, solar, and batteries we chose 1980 as earliest years. Since biomass (with solid biomass fuel), combustion, and hydro power plants are older technologies, we allowed earlier installation years. We also checked that rated powers are in acceptable range, ie. for biomass (0MW - 150MW), for combustion (0GW - 2GW), for hydro (0GW - 1.5GW), for solar (0MW - 500MW), for storages including pumped hydroelectric energy storages (0MW - 800MW), and for wind (0MW - 22MW). These values need to be adapted over time. For wind turbines, we also checked that the hub height is larger than the rotor radius, and for PV systems we checked that balcony PV systems have a rated power $P < 1$kW.

\end{itemize}

In total we have defined 90 data unit tests for the MaStR dataset.

\section{Results} \label{sec:results}
We ran the data pipeline and tests as described in Section \ref{sec:method} on the MaStR dataset obtained at 2024-03-12. All units that failed at least one test can be browsed online at \href{https://marktstammdaten.kotthoff.dev}{marktstammdaten.kotthoff.dev}. If the data is needed for further processing, it can be obtained from there as csv or json. 
\subsection{Basic Tests}
As expected, most of the basic tests do not find errors. This shows that basic integrity of the MaStR dataset is granted. Basic expectations, like the existence of unique unit IDs are fulfilled. We only found two biomass units that do not have a municipality ID. Moreover, 31 hydropower units, roughly 30 solar and storage units that are not allocated at a municipality (and hence have no municipality ID and no zip code). We also found 6 storages where the inverter power was larger than the net power. 
\subsection{System size tests}
\begin{figure*}[t]
    \centering
    \begin{minipage}[c]{0.75\textwidth}
        \includegraphics[width=\textwidth]{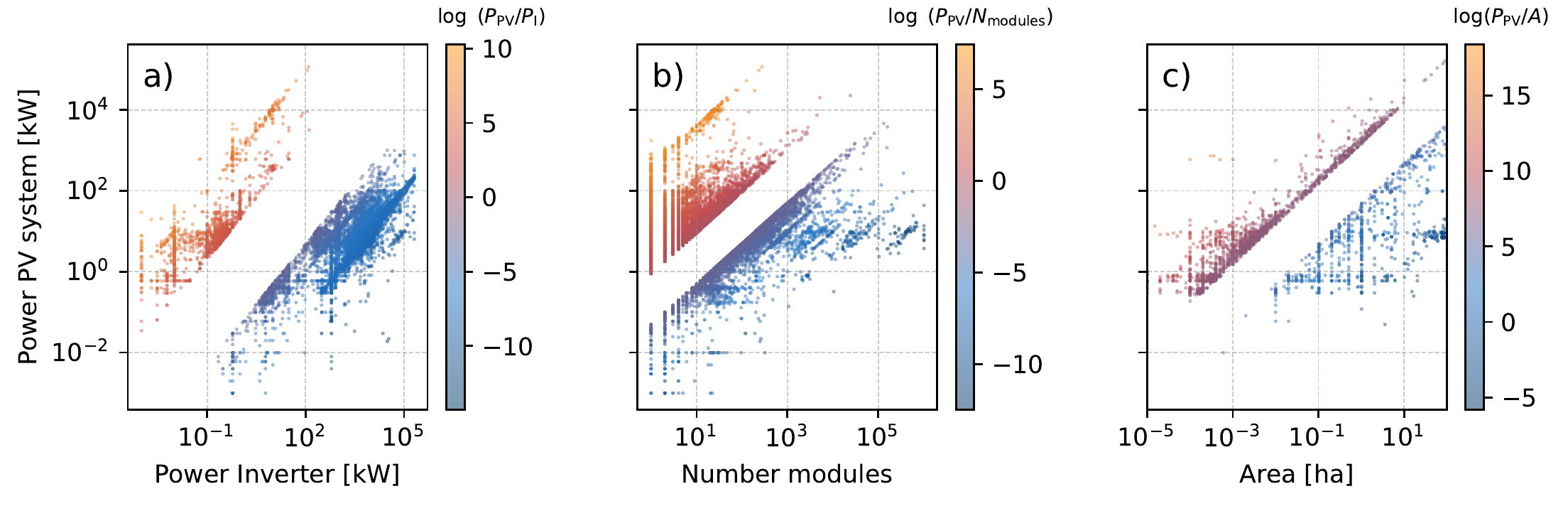}
    \end{minipage}%
    \hfill
    \begin{minipage}[c]{0.25\textwidth}
    \vspace*{5pt}
        \includegraphics[width=\textwidth]{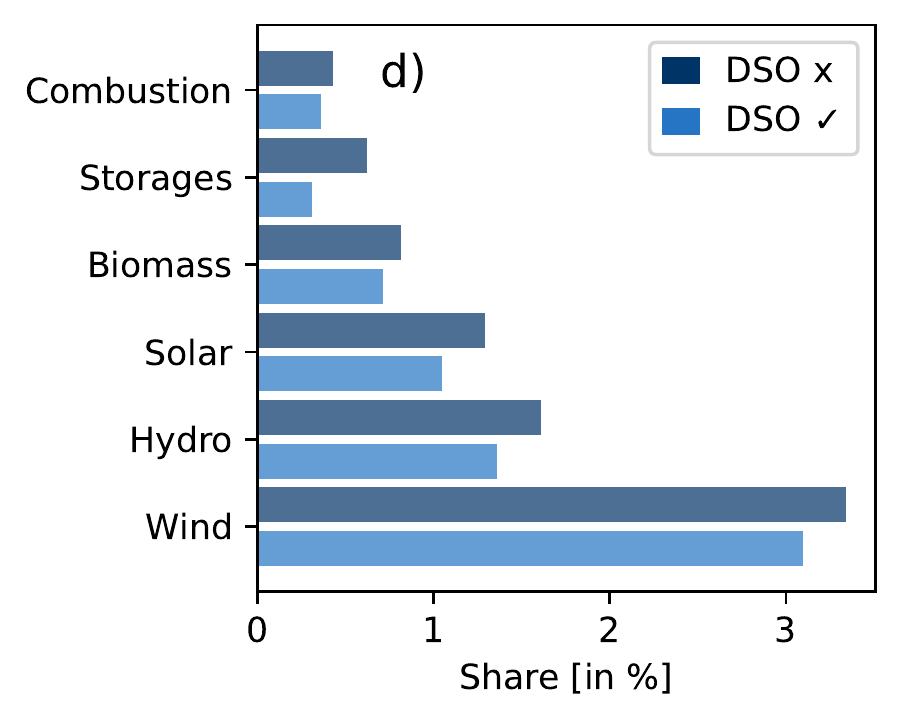} 
    \end{minipage}
    \caption{Results of the System size and Location Tests. The white bands in the three plots represent the ranges of allowed values. In the scatter plots a)-c) each dot represents one PV system. In a) the power of the PV modules is compared to the power of the inverter, in b) the power of the PV modules is comapared to the number of modules, and in c) the power of ground-mounted PV systems is compared to the area needed for the installation. Colors represent the log value of the fraction of y-axis and x-axis. In d) the share of units that have mismatching coordinates and districts is shown, where dark blue represents all units of the MaStR (including units that are not yet validated by DSOs).}
    \label{fig:results} 
    \Description{Three scatter plots are shown. In the first plot a), the Power of PV systems is plotted on the y-axis ranging from $10^{-2}$kW to $10^5$kW. On the x-axis the Power of assigned converters is shown, ranging from $10^{-2}$kW to $10^5$kW. In the second plot b), the Power of PV systems is plotted on the y-axis ranging from $10^{-2}$kW to $10^5$kW. On the x-axis the Number of modules is shown, ranging from $10^{0}$ to $10^6$. In the third plot c), the Power of PV systems is plotted on the y-axis ranging from $10^{-2}$kW to $10^5$kW. On the x-axis the Area in ha is shown, ranging from $10^{-5}$ha to $10^2$ha.}
\end{figure*}

The second error category is about system size. We compared values from the dataset that correlate with the system size and compared it to the rated power. In Fig. \ref{fig:results} a) we see that for 23,251 PV system, each represented by one dot in the scatter plot, the installed power of the PV modules and the power of the inverter differ with a factor 20 or more. This represents a share of roughly 0.6\% of PV units. Especially the higher densities of points, where the units' power differ of a factor of $10^3$ shows a typical error, where the physical quantities W, kW, and MW are mixed up. In b) we compared the installed power of PV systems with the number of units, where we accepted a range of 50W - 700W per module. We selected these boundaries with an additional buffer to ensure that any raised errors are genuine. As result we see that for 43,733 (representing 1.12 \%) of PV units the number of modules do not match with the installed PV power, where a large number of units are registered with only one module, possibly as a placeholder number. 
In c) we compared the installed power of ground-mounted PV systems with the area they needed. Knowing the are of ground-mounted PV systems is important for understanding the conflict between using arable area for agriculture or for electricity production. We see that the large share of 21.3\% (3758 units) do not lie within our accepted range of 0.05MW/ha - 1.5MW/ha.  

For wind turbines, the rated power scales quadratically with the rotor diameter $P \backsim r^2$. From \cite{gonzalez2016_windrotor} we took accepted ratios to be $160W/m^2$ - $700W/m^2$ and found 302 units (0.8$ \% $) that do not lie in this range. This is a rather small share, and we also saw that most erroneous wind turbines are small turbines with rated powers $\ll$ 1MW. For storages, we only found 770 storages where the inverter power and the power of the storage differ with a factor 20 or more.

For the other technologies, we couldn't find test cases that only use data from within the MaStR to verify the system sizes.

\subsection{Location Tests}
When testing if the coordinates of units lie within their specified districts, we found that especially for wind turbines, a high share of more than 3\% of units have non matching coordinates and districts (see Figure \ref{fig:results}d). This corresponds to an accumulated installed power of more than 2 GW. 
Here we also see a concentration of location errors, where roughly half of the wind turbines with erroneous locations stem from only five districts.
For all other technologies, the share of units with wrong location information is smaller (between 0.5 \% and 2 \%). We compared the amount of location information of the whole dataset with the subset of entries in the dataset that were already corrected by the DSO and find that the data quality is better after the DSO check, but many location errors passed these checks unnoticed.

 One possible reason why those erroneous entries exist is the continuous change of geographic areas of districts and municipalities. In the MaStR, districts and municipalities occur even though they do not exist anymore because they were legally integrated into other districts or renamed. The entries within the MaStR are not automatically updated if above legal change occurs. However, not all erroneous entries can be traced back to this issue and some entries are simply wrong. We saw several wind farms with turbines allocated to different districts, even though next district borders were far away. In the appendix \ref{sec:appendix_distance}, we further evaluate the distance between unit coordinates and their districts. 

We also tested coordinates against the specified municipalities of units and found about 10 GW of wind turbines, 0.3 GW of hydro power plants, 2.9 GW of solar power plants, 0.2 GW of biomass plants, 0.2 GWh of storages, and 4.3 GW of combustion power plants that had wrong location information regarding their registered municipalities.

\subsection{Technology specific Tests}
When testing the rated power of units (Test ID 12), we did not find units lying outside of the accepted ranges. For the installation years, we found 287 storages and 1169 PV systems to have installation dates earlier than 1980. For example, the MaStR contains battery storages with a rated power of more than 1MW for the installation year 1923. However, the accumulated power of all units with unplausible installation years is only of the order of several MW. 

A rather unusual error comes up when testing the hub height of wind turbines. Here we found 4 turbines with a hub height smaller than the rotor radius, resulting in either broken turbine blades or a well-plowed field. Finally, we also tested the rated capacity of balcony-PV systems. In Germany, they fall under an own category of units with a simplified installation and registration, but they are only allowed to have a net power < 600W, which is planned to be increased to 800W within the year 2024. We first tested all units with unit type "balcony PV" to have a rated power of less than 1kW (with a small error tolerance of 200W). Then we also tested that units that have the word "balcony" in their name should not have more than 5kW. We found about 9822 units failing one or both of these tests.

\subsection{Updating results regularly}
The MaStR dataset is updated once per day. Every update introduces new units that might contain erroneous data. Every update can also fix errors in existing entries. At the moment of reading this paper, the results section is already outdated. To overcome this issue, we decided to package the whole data pipeline and testing in one shell script, so that it can be triggered regularly. By using the framework datasette \cite{Willis2024_Datasette}, it is then possible to visualize and publish the test results in an interactive website. All relevant plots of this results section are part of this dashboard and can be updated regularly. We host the current version of the dashboard at \url{https://marktstammdaten.kotthoff.dev/-/dashboards/verify-marktstammdaten}.  

\section{Discussion} \label{sec:discu}
The MaStR is a rich and valuable dataset, both for the control and operation of the energy system as well as for research. This is underlined by the increasing number of publications that utilised the MaStR dataset, as we have shown in our literature review. We only considered publications in scientific journals. However, we assume that the utilization of the registry in applied science\footnote{For example visualization tools like \url{https://www.energy-charts.info/charts/power/chart.htm?l=en&c=DE} or regional case studies like \url{https://web.archive.org/web/20221108120815/https://opus.hs-osnabrueck.de/frontdoor/deliver/index/docId/3628/file/Energieversorgung.pdf}} is also of high relevance. Because of this, we're convinced that monitoring the data quality and highlighting possible shortcomings of this core dataset from the German energy system is an important contribution.

Regarding our results, we saw that the \textbf{Basic Tests} do not uncover crucial errors in the MaStR dataset, as most of them find none or few errors. Their main purpose comes when running the data and testing pipeline regularly to detect very fundamental mistakes (like non unique unit IDs). This is also true for the \textbf{Technology specific Tests}, especially when checking that rated powers lie in acceptable ranges. Testing hub heights of wind turbines or rated powers of balcony PV system shows one important aspect of data tests: For some tests, domain knowledge is crucial. A collaborative approach that allows experts from various energy domains to easily define and implement data tests would significantly enhance the data quality of the MaStR, benefiting the entire community.

The \textbf{Location Tests} resulted in a mismatch between unit coordinates and assigned districts and municipalities for a substantial amount of units. Thus, researchers intending to use regional sub-samples of the registry should be cautious when selecting and aggregating data. It would be at least awkward to present large renewable energy plants in regional case studies which do not exist there.

The \textbf{System Size Tests} were an effective way of flagging erroneous entries. However they cannot be easily used to correct the data in the MaStR, as it is not clear which of the two properties is correct. For example, when seeing that an inverter and the modules power differ with a factor of $10^3$, one of them is wrong for sure, but we cannot say which one. As a small contribution to also improve the data quality of the MaStR, we added a sub page \footnote{Visit \url{https://marktstammdaten.kotthoff.dev/-/dashboards/erronous-entries-district}} where system size errors can be obtained on a district level. We then disseminated this site over a Mail distribution list for the german energy domain. We hope that some DSOs are concerned about the data quality of the MaStR and try to correct the entries within their districts. As DSOs usually have the technical reports of both the installed modules and the inverter, this should be feasible.

To foster transparency and repeatability of results, we make our code openly available. The described dasboards can provide valuable insights into data quality of the MaStR. The pipeline and dashboard creation scripts are uploaded on github \cite{githubrepo}.

\subsection{Limitations of our work}

Our analysis is subject to limitations that we want to state clearly:

\begin{itemize}
    \item To design elaborated test cases, domain knowledge is needed. Our knowledge helped us designing many test cases, but we are convinced that tests are missing. A process for different domain experts to collaboratively add data tests would be beneficial for the whole community.
    
    \item We only validate data from within the MaStR, thus our results only apply for the dataset itself. We also do not crosscheck the content with other data sources. One further step could be to compare units registered in the MaStR with aerial images using deep learning based segmentation algorithms, as in \cite{Schulz2021_DetEEktor, Mayer2022_3D}.
    
    \item We do not correct data in the original MaStR dataset. This would be an important step, but requires the collaboration of more stakeholders, including the Federal Network Agency and DSOs.
\end{itemize}

\section{Conclusion} \label{sec:conclu}
As energy systems grow more complex, both energy research and operation needs to be based on detailed and reliable data to provide sound insights. 
Our contribution to this goal is a first analysis of the data quality of the MaStR as the central power and energy unit registry of Germany. 

We start by emphasizing the urgency to validate the dataset by investigating its relevance in scientific literature. We find an increasing number of studies that use the MaStR data since its first release in January 2019. 

In our opinion, the MaStR dataset is crucial for controlling and understanding the transforming energy system. It contains a huge amount of valuable and high-quality information, bearing in mind the identified issues with some dimensions of data quality. In the future, we recommend to increase the validation effort for the MaStR data with the help of automated techniques, together with scientific and public stakeholders.

In conclusion, we see this work as a first step towards a continuous validation process of the MaStR dataset. Validation efforts of the MaStR are most-likely already performed by many of its users. However, we think that results and methods should be shared to avoid redundant work and save resources which could be used better in producing valuable research of sustainable energy systems. Researchers using the MaStR dataset should be aware of the existing shortcomings and our identified errors. Wrong locations, for example, are of special relevance in the case of regional potential analyses, where it is important that the renewable power plants used in the studies were actually built within this region. 
With this paper and our data testing pipeline we have made a first contribution to an ongoing and open validation process.

\begin{acks}
Blinded for review process.
\end{acks}

\appendix

\section{Appendix: Dataset Description}
\label{appendix:tables}
To give a clear idea of the content that we tested in the MaStR dataset, we provide table \ref{tab:appendix}. It shows all columns that we have analyzed, together with an example entry and an indication of whether this column is apparent in the different technology tables. An x denotes that the column does not exist for this table, a number < 100 denotes the share of entries that are not null.

\begin{table*}[h!]
    \include{appendix/all_techs}
    \caption{Overview of selected columns with example and completeness for each table, where completeness is defined as fraction of number entries that are not null to the total number of entries (For example: 97\% of units in the biomass table have a "unit owner mastr id").  An x marks columns that do not exist for this table.}
    \label{tab:appendix}
\end{table*}

\section{Appendix: Distance of wrongly allocated units}
\label{sec:appendix_distance}
\begin{figure}[h!]
    \centering
    \includegraphics[width=\columnwidth]{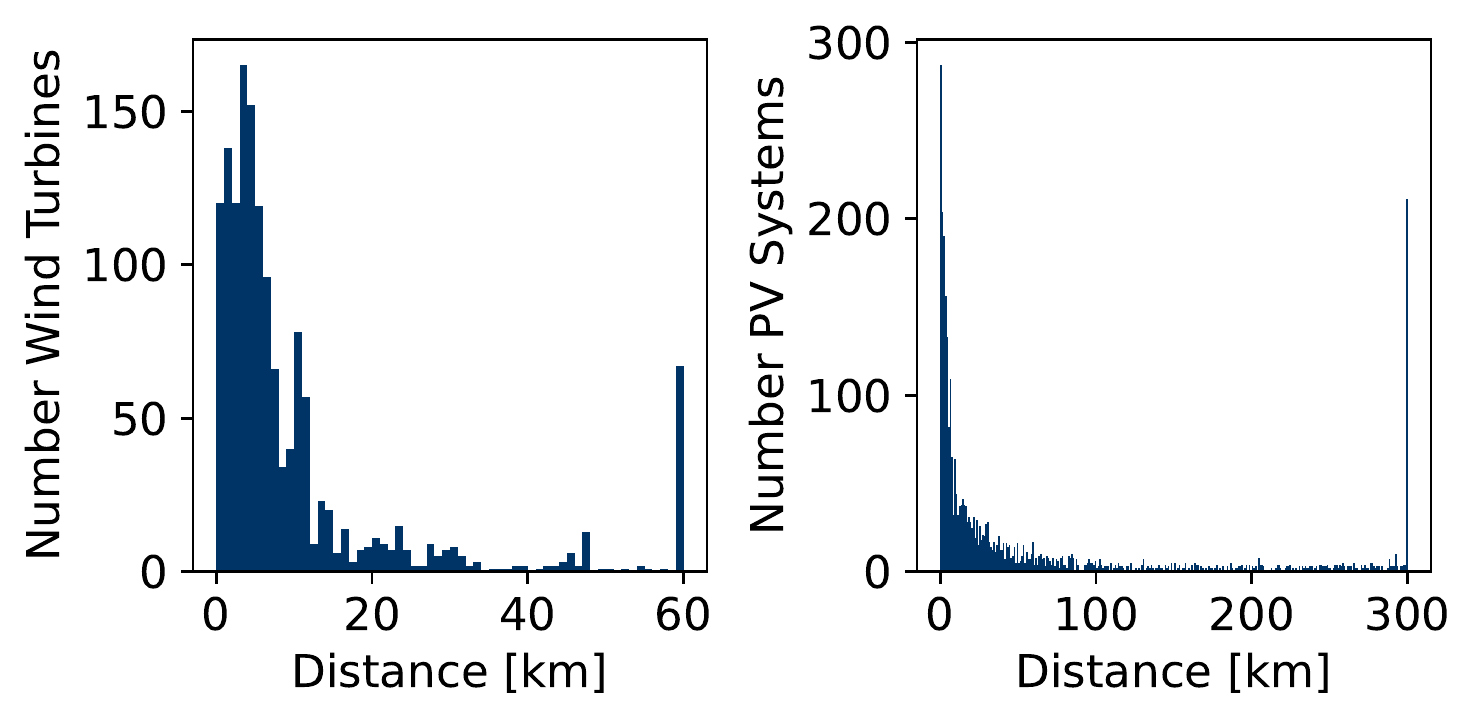}
    \caption{Number of units where the coordinates do not lie within the district, plotted over the distance of the coordinates to the district boundaries. For wind turbines (left) we see that most turbines are close to their district with distance < 40km. For PV systems, the distance to the district can be larger. The last bar in both plots is relatively large, since we used the aggregated value of all wind turbines that have a distance larger than 60km (solar larger than 300km).}
    \label{fig:Appendix_Distance}
    \Description{}
\end{figure}
As described in section \ref{sec:results}, the coordinates and assigned districts did not match for all units. Especially for wind and PV systems, we found erroneous entries. To see how far coordinates and districts lie apart, we calculated the distance between coordinates and district boundaries and plotted the results as a histogram in Figure \ref{fig:Appendix_Distance}. We see that most of the units, where the coordinates and districts do not match, lie relatively close to the district boundaries. This is true for both wind turbines and PV systems. Only for units with a distance of several km we assume that the reason for the errors cannot be due to changes to the district boundaries over time.

\bibliographystyle{ACM-Reference-Format}
\bibliography{mastr.bib}    
\end{document}

%% file: figures/table_testcases.tex
\centering

\begin{tabular}{|c|c||c|c|c|c|c|c|}

\hline
\rotatebox{270}{Test ID} & description & \rotatebox{270}{biomass} & \rotatebox{270}{combustion  } & \rotatebox{270}{hydro} & \rotatebox{270}{solar} & \rotatebox{270}{storages} & \rotatebox{270}{wind} \\
\hline
1 & Check for null values: unit ID, municipality ID, operating status, power  & \checkmark & \checkmark & \checkmark & \checkmark & \checkmark & \checkmark \\
2 & Check that unit IDs are unique  & \checkmark & \checkmark & \checkmark & \checkmark & \checkmark & \checkmark \\
3 &$P_{gross}$ $\geq$ $P_{net}$&  &  &  & \checkmark & \checkmark &  \\
4 &$P_{inverter}$ $\geq$ $P_{net}$&  &  &  & \checkmark & \checkmark &  \\
5 & Check that values match regex: unit ID, municipality ID, zip code  & \checkmark & \checkmark & \checkmark & \checkmark & \checkmark & \checkmark \\
\hline
6 & Check that $P_{gross}$ / number modules is in accepted range  &  &  &  & \checkmark &  &  \\
7 & Check that $P_{gross}$ / $P_{inverter}$ is in accepted range &  &  &  & \checkmark & \checkmark &  \\
8 & Check that $P_{gross}$ / area is in accepted range (only ground-mounted PV) &  &  &  & \checkmark &  &  \\
9 & Check that rated power matches with rotor diameter  &  &  &  &  &  & \checkmark \\
\hline
10 & Check that the coordinates lie in the district  & \checkmark & \checkmark & \checkmark & \checkmark & \checkmark & \checkmark \\
11 & Check that the coordinates lie in the municipality  & \checkmark & \checkmark & \checkmark & \checkmark & \checkmark & \checkmark \\
\hline

12 & Check that the installed power is reasonable  & \checkmark & \checkmark & \checkmark & \checkmark & \checkmark & \checkmark \\
13 & Check that the installation year is reasonable  & \checkmark & \checkmark & \checkmark & \checkmark & \checkmark & \checkmark \\


14 & Check that the hub height matches with rotor diameter  &  &  &  &  &  & \checkmark \\
15 & Check that balcony PV systems have a small installed capacity  &  &  &  & \checkmark &  &  \\

\hline
\end{tabular}

%% file: appendix/all_techs.tex
\begin{tabular}{llllllll}
\toprule
column & example & $C_{biomass}$ & $C_{combustion}$ & $C_{hydro}$ & $C_{solar}$ & $C_{storages}$ & $C_{wind}$ \\
\midrule
unit owner mastr id & ABR989393706204 & 97 & 95 & 99 & 100 & 100 & 97 \\
mastr id & SEE900002935310 & 100 & 100 & 100 & 100 & 100 & 100 \\
operating status & In Betrieb & 100 & 100 & 100 & 100 & 100 & 100 \\
grid operator inspection & 1 & 100 & 100 & 100 & 100 & 100 & 100 \\

commissioning date & 2001-12-21 & 99 & 99 & 100 & 98 & 99 & 91 \\
planned commissioning date & 2024-12-31 & 1 & 1 & 0 & 1 & 1 & 9 \\
installation year & 2017 & 100 & 100 & 100 & 100 & 100 & 100 \\
download date & 2024-03-12 & 100 & 100 & 100 & 100 & 100 & 100 \\

zip code & 17291 & 100 & 100 & 100 & 100 & 100 & 95 \\
municipality & Bad Wünnenberg & 100 & 100 & 100 & 100 & 100 & 95 \\
district & Nordfriesland & 100 & 100 & 100 & 100 & 100 & 95 \\
coordinate & 48.1748, 11.5961 & 96 & 26 & 55 & 5 & 0 & 97 \\

power gross & 5 & x & x & x & 100 & 100 & x \\
power inverter & 10 & x & x & x & 100 & 100 & x \\
power net & 5 & x & x & x & 100 & 100 & x \\
power & 2000 & 100 & 100 & 100 & x & x & 100 \\

combustion technology & Verbrennungsmotor & 100 & x & x & x & x & x \\
fuel type & Gasförmige Biomasse & 100 & x & x & x & x & x \\

energy carrier & Erdgas & x & 100 & x & x & x & x \\

type of inflow & Flusskraftwerk & x & x & 87 & x & x & x \\
plant type & Laufwasseranlage & x & x & 100 & x & x & x \\

combination with storage & Kein Stromspeicher & x & x & x & 98 & x & x \\
number of modules & 8 & x & x & x & 98 & x & x \\
orientation & Süd & x & x & x & 98 & x & x \\
orientation secondary & West & x & x & x & 21 & x & x \\
unit type & Freifläche & x & x & x & 100 & x & x \\

storage capacity & 10 & x & x & x & x & 100 & x \\
battery technology & Lithium-Batterie & x & x & x & x & 100 & x \\

technology & Horizontalläufer & x & x & x & x & x & 100 \\
type description & E-70 E4 & x & x & x & x & x & 99 \\
manufacturer & ENERCON GmbH & x & x & x & x & x & 99 \\
position & Windkraft an Land & x & x & x & x & x & 100 \\
hub height & 65 & x & x & x & x & x & 98 \\
rotor diameter & 82 & x & x & x & x & x & 99 \\
\bottomrule
\end{tabular}